\documentclass{mem}
\pdfoutput=1
\usepackage{natbib}\usepackage{txfonts}\usepackage{balance}
\usepackage{graphicx}
\usepackage[a4paper]{hyperref}
\usepackage{url}

\def\makeheadbox{
\parbox[t]{\textwidth}{%
  Proceedings 25th NSO Workshop \newline
    Eds.~A.~Tritschler, K.~Reardon, H.~Uitenbroek \newline
      Mem. S.A.It., in press}
      \par\vspace*{-12ex}\mbox{}
}

\begin{document}

\title{High frequency waves in the solar atmosphere?}

\subtitle{}

\author{
B. \,Fleck\inst{1}
\and T. \,Straus\inst{2} 
\and M. \,Carlsson\inst{3}
\and S. \ M.\,Jefferies\inst{4}
\and G. \,Severino\inst{2}
\and T. D. \,Tarbell\inst{5} }

\offprints{B. Fleck}
 
\institute{
ESA Science Operations Department, 
c/o NASA Goddard Space Flight Center, Mailcode 671.1, Greenbelt, MD 20771, USA, 
\email{bfleck@esa.nascom.nasa.gov}
\and
INAF / Osservatorio Astronomico di Capodimonte
Via Moiariello 16
80131 Napoli, Italy
\and
Institute of Theoretical Astrophysics, University of Oslo, Norway
\and
Institute for Astronomy, University of Hawaii, HI 96768, USA
\and
Lockheed Martin Solar and Astrophysics Laboratory, Palo Alto, CA 94304, USA 
}

\authorrunning{Fleck et al.}

\titlerunning{High frequency waves in the solar atmosphere?}

\abstract{
The present study addresses the following questions: How 
representative of the actual velocities in the solar atmosphere 
are the Doppler shifts of spectral lines? How reliable is the 
velocity signal derived from narrowband filtergrams? How well 
defined is the height of the measured Doppler signal? Why do 
phase difference spectra always pull to 0$^\circ$ phase lag 
at high frequencies? Can we actually observe high frequency 
waves ($P\le 70$~s)? What is the atmospheric MTF of high frequency waves? 
How reliably can we determine the energy flux of high frequency 
waves? We address these questions by comparing observations obtained 
with Hinode/NFI with results from two 3D numerical simulations 
(Oslo Stagger and CO$^5$BOLD). Our results suggest that the 
observed high frequency Doppler velocity signal is caused by 
rapid height variations of the velocity response function in 
an atmosphere with strong velocity gradients and cannot be 
interpreted as evidence of propagating high frequency acoustic 
waves. Estimates of the energy flux of high frequency waves 
should be treated with caution, in particular those that apply 
atmospheric MTF corrections. 
\keywords{Waves -- Line: formation -- Sun: chromosphere -- Sun: oscillations -- Sun: photosphere }
}

\maketitle

In the present study we compare Mg~b$_2$ and Na~D$_1$ Dopplergram
time series with results from two 3-D simulations. The observations
were obtained with the narrowband filter (NFI) on Hinode. The
Mg\,b$_2$ series was obtained on 2009/01/11. It is about 2 hours long
and has a cycle time of 32\,s. Each cycle comprises 4 filtergrams,
taken at $\pm 68$~m\AA\ (``core'') and $\pm 188$~m\AA\ (``wing'')
from the line center position, respectively. The Na~D$_1$ series was
obtained on 2009/01/07, with same duration and cycle time and
wavelength offsets of $\pm 80$~m\AA\ and $\pm 168$~m\AA,
respectively. After coalignment and interpolation to a common time
frame, velocity proxies have been calculated by
\mbox{$S_v=(R-B)/(R+B)$}, where $R$ and $B$ denotes the measured red
and blue wing (core) intensities.
\begin{figure*}[]
\resizebox{\hsize}{!}{
\includegraphics[clip=true]{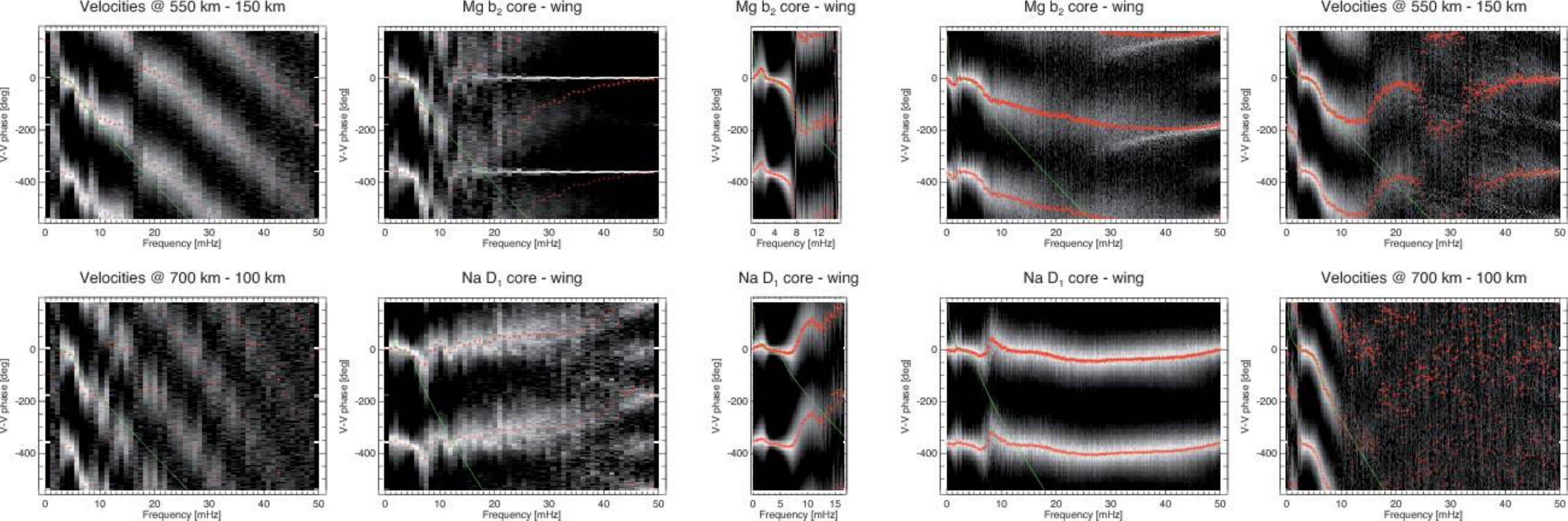}}
\caption{\footnotesize
Comparison between observations and simulations of the Mg~b$_2$ 
(upper row) and Na~D$_1$ (lower row) lines:
The two smaller panels at the center display the spatially averaged 
phase difference scatter plots obtained from observations 
with Hinode/NFI. The deviation from the expected behavior 
of linear sound waves (green line) is evident. The outermost 
panels on either side show phase difference spectra between actual velocities 
(heights indicated in diagrams), with the Oslo-Stagger model 
on the left and CO$^5$BOLD model on the right. The inner 
panels show the phase difference between core and wing 
Doppler shifts from the simulations. }
\label{Fleck-fig:phase}
\end{figure*}
\begin{figure*}[]
\resizebox{\hsize}{!}{\includegraphics[clip=true]{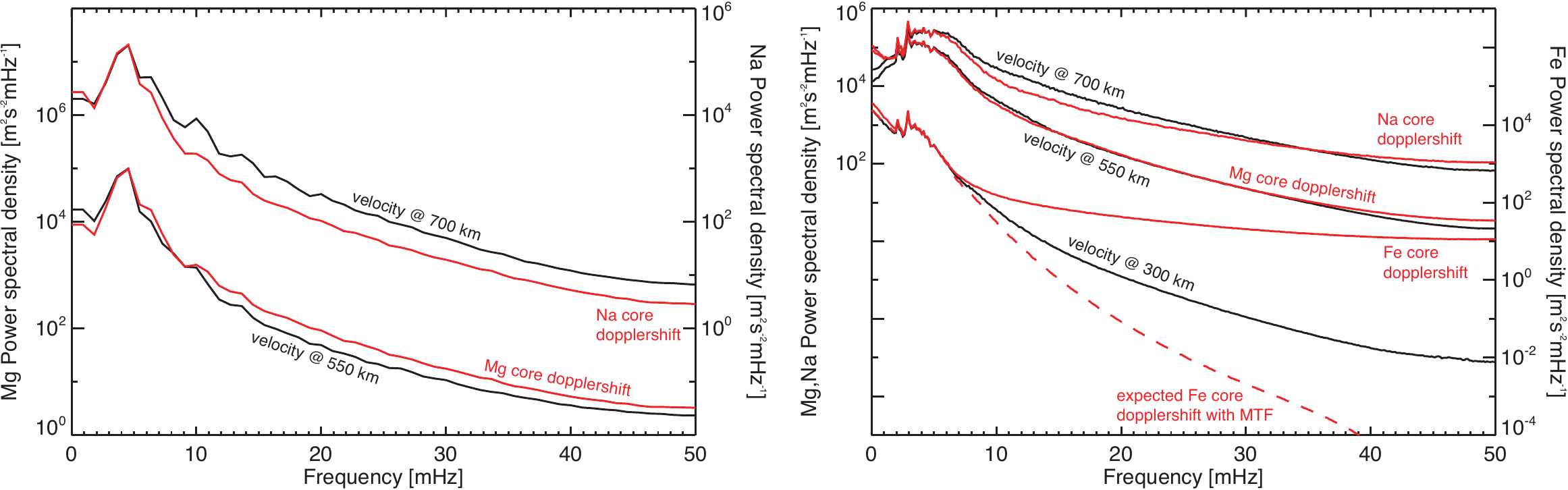}}
\caption{\footnotesize
Power spectra of the simulated Hinode/NFI signals and of the 
actual velocities at corresponding heights in the simulations. 
The Oslo-Stagger code results are shown in the left panel, 
the CO$^5$BOLD model results are on the right. The expected 
MTF for Fe~6301 is taken from \cite{2008ESPM...12.2.39F}. 
Surprisingly, none of the power spectra of the simulated 
observations shows such a behavior. Instead, the ``observed'' 
power is comparable to or higher than the power of the actual velocities. }
\label{Fleck-fig:power}
\end{figure*}
\begin{figure*}
\centering
\resizebox{\hsize}{!}{
\includegraphics[clip=true]{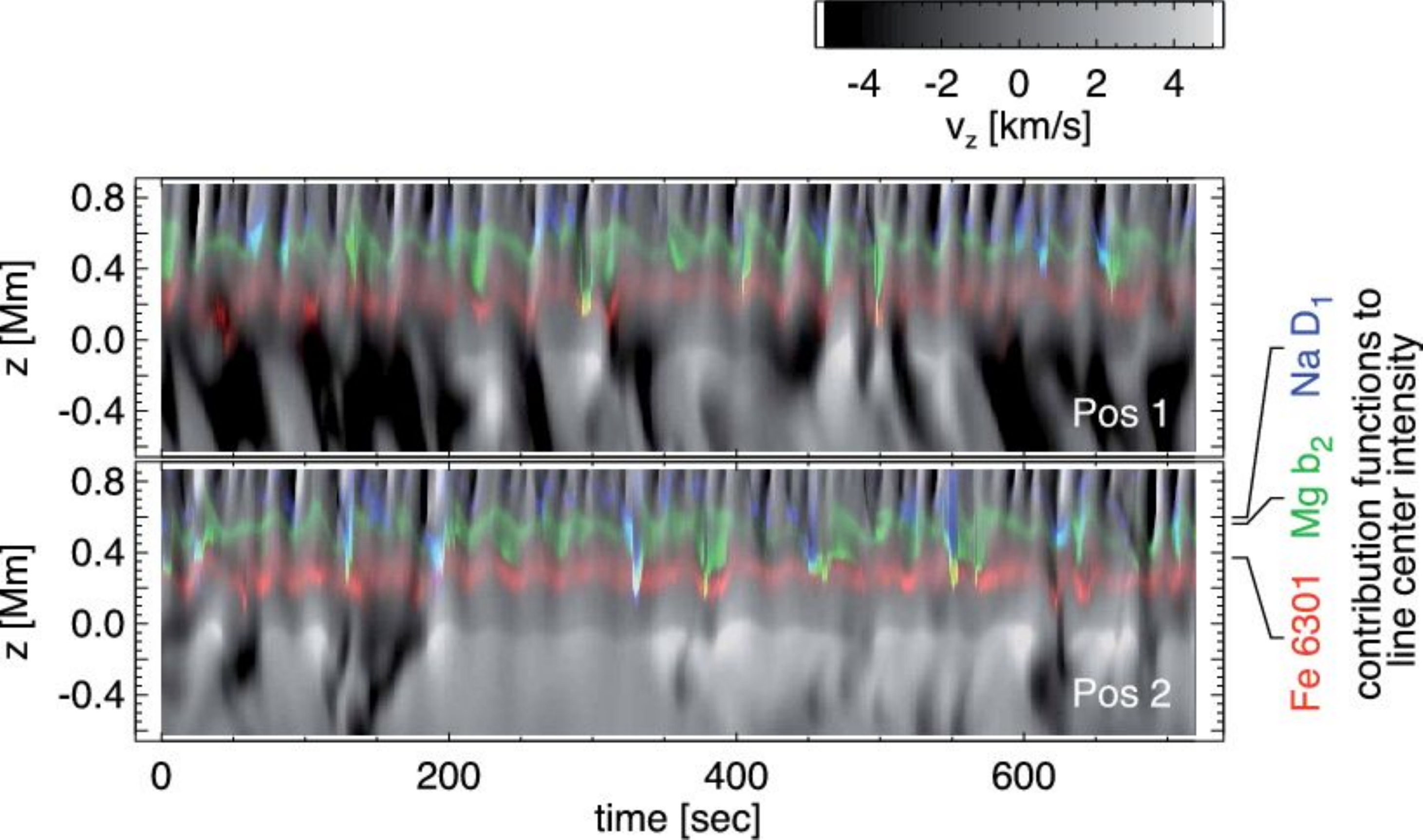}}
\caption{\footnotesize
$z-t$ slices of the CO$^5$BOLD simulation with contribution functions of 
Fe~6301 (red), Mg~b$_2$ (green), and Na~D$_1$ (blue). 
Note the rapid and significant height variations 
of the contribution functions. }
\label{Fleck-fig:Contrib}
\end{figure*}
\begin{figure*}[]
\centering
\resizebox{0.9\hsize}{!}{
\includegraphics[clip=true]{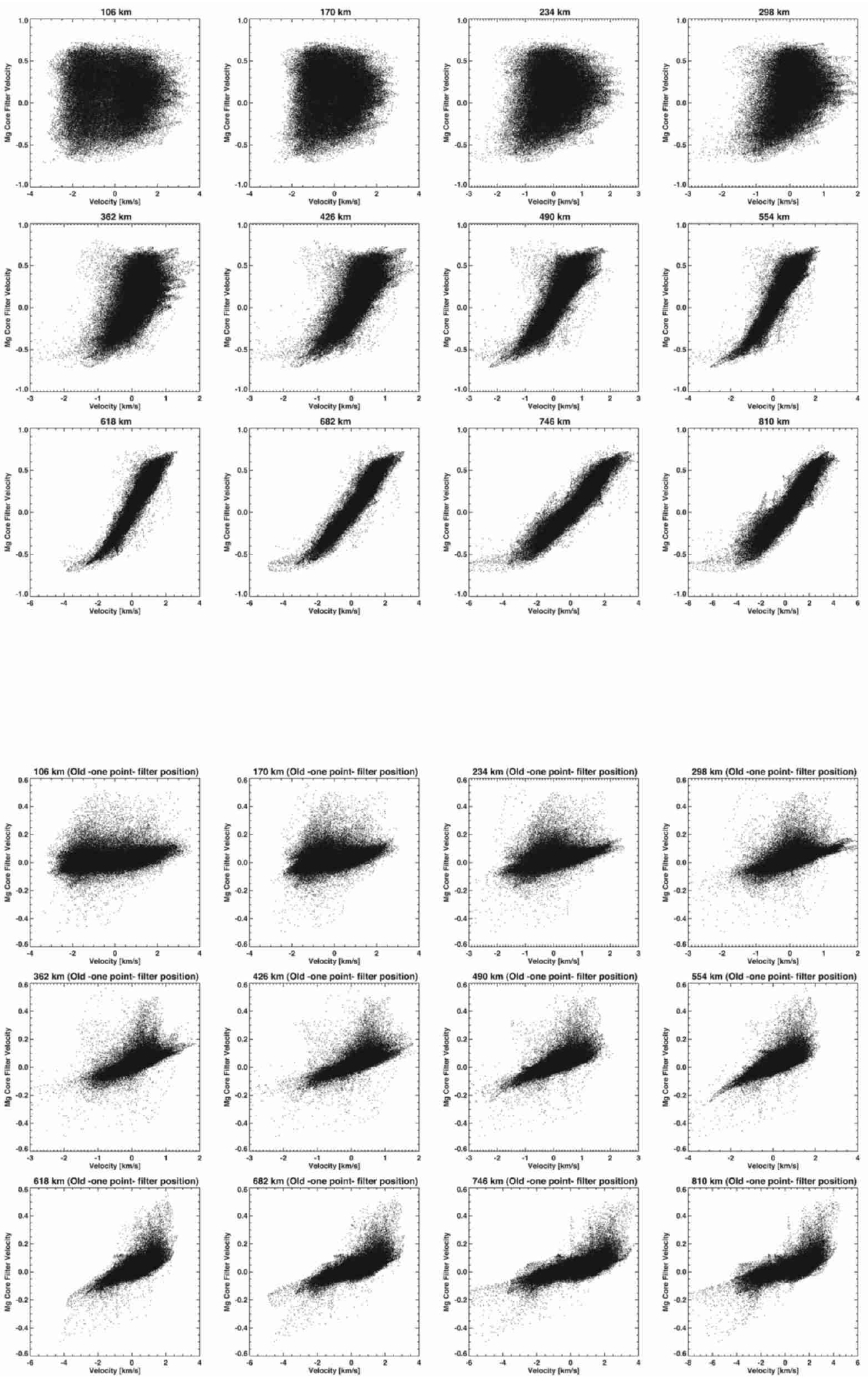}}
\caption{\footnotesize
Scatter plots of simulated Mg ``core''  Dopplergram signals versus 
actual velocities of the Oslo-Stagger model at various heights. 
The upper set of panels shows results for the filter settings 
of the four-point measurement used in this work, with the ``core'' 
measurements taken at at $\pm 68$~m\AA\ from line center. 
The lower panels show corresponding results for the "old" two-point 
settings at $\pm 113$~m\AA\ from the line core. The latter 
setting clearly yields inferior velocity measurements, as it 
mixes photospheric and chromospheric signals. }
\label{Fleck-fig:scat}
\end{figure*}

The two simulations we use were computed with the Oslo-Stagger
code~\citep{2007ASPC..368..107H} and the CO$^5$BOLD
code~\citep{2004A&A...414.1121W,2008ApJ...681L.125S}. They give access
to both the actual velocities at given heights in the atmosphere as
well as ``observed'' Doppler shifts in simulated line profiles. The
Oslo-Stagger simulation covers 20 minutes of solar time and extends
from the subphotospheric convection zone up to the corona. In this
simulation, the Mg~b$_2$ and Na~D$_1$ lines have been calculated in
1D NLTE. The CO$^5$BOLD simulation covers 2 hours of solar time and
has the upper boundary at approximately 900~km above the base of the
photosphere. The line profiles of the Fe~6301, Mg~b$_2$, and
Na~D$_1$ lines have been calculated in LTE. For this simulation we
also have the full contribution functions to the emergent intensity at
the minimum of the line profiles in each spatial point and for each
time step. Furthermore, the velocity response function has been
calculated for each spatial point for the first snapshot of the time
series. The resulting estimates of the average formation heights are
150~km and 550~km for the wing and core Doppler shifts in the
Mg~b$_2$ line, and 100~km and 700~km for the wing and core signals
in the Na~D$_1$ line, respectively.

The phase difference spectra of the actual velocities in the
Oslo-Stagger simulations show the expected behavior of linear wave
theory~\citep{1966AnAp...29...55S}, with a linear phase increase up to
the Nyquist frequency (Fig.~\ref{Fleck-fig:phase}). The corresponding
CO$^5$BOLD spectra show good agreement up to about 10~mHz. At higher
frequencies, they do not follow the expected behavior but reveal
several $\pm180^\circ$ phase jumps, suggesting wave
interference. Inspection of time-lapse sequences of x-z cuts of the
CO$^5$BOLD cube indeed suggests wave reflection near the upper
boundary and at the steepening shock fronts in the chromospheric
layers. This aspect of the CO$^5$BOLD simulations requires further
investigations. 

Turning to the phase differences between the simulated Doppler shifts
(inner panels of Fig.~\ref{Fleck-fig:phase}), they are --- in both
simulations --- markedly different from those of the corresponding
actual atmospheric velocities. Up to 10~mHz they are similar to the
observed spectra. This suggests that the unexpected shape of the
observed phase difference spectra at high frequencies is caused by
line formation effects of the Dopplergram signal rather than by unexpected
propagation properties of high frequency sound waves. One is tempted
to attribute this to the suppression of high-frequency waves by the
atmospheric MTF due to the extended width of the velocity response
functions~\citep[cf.][]{1986ApJ...310..912K}. 

To check this hypothesis we compared power spectra of the simulated
Doppler shifts with those of the actual velocities at the target
heights in the simulations (Fig.~\ref{Fleck-fig:power}). As the
Dopplergram signals have not been calibrated, we normalized them to the
total power of the actual velocities in the frequency range from 3 to
6~mHz. Surprisingly, the power spectra of the Doppler shifts do not
reveal the expected steep falloff at high frequencies. Instead, the
power at high frequencies is comparable to the power of the actual
velocities. In the case of the Mg core Doppler shift the power at high
frequencies is even higher than that of the actual velocities. This
discrepancy is most evident in the case of the Fe~6301 line, for
which the high-frequency power of the Doppler shifts is orders of
magnitude higher than the power of the actual velocities, although the
expected MTF should reduce the power by 3 orders of magnitude at
50~mHz \citep{2008ESPM...12.2.39F}. The reason for this becomes clear
upon inspection of the contribution functions, which shows rapid and
considerable height variations (Fig.~\ref{Fleck-fig:Contrib}). It
appears that the observed high frequency Doppler signal is \emph{not}
due to propagating high frequency acoustic waves, but due to fast and
significant height variations of the velocity response function in a
dynamic atmosphere with strong vertical velocity gradients.

Scatter plots of ``observed'' ``core'' Doppler shifts versus actual
velocities show a good correlation at chromospheric heights (upper
panel of Fig.~\ref{Fleck-fig:scat}), suggesting that the dominant
velocity component can be measured reasonably well with simple
Dopplergrams and that the Mg ``core'' signal measured at $\pm
68~$m\AA\  from line center is indeed a useful measure of the
velocities in the lower chromosphere. A previous set up of NFI for
two-point measurements in Mg~b$_2$ at $\pm 113$~m\AA\ from the line
core (i.e. close to the knee between the Doppler core and the damping
wings) shows a bifurcated distribution with much reduced correlation
(see lower panels in Fig.~\ref{Fleck-fig:scat}). At that filter
position, the measured intensities are a complex mixture of
photospheric and chromospheric signal. Work is in progress to better
calibrate the NFI Dopplergrams and to determine the optimum filter
position.

We conclude that previous claims of the detection of high frequency
waves ($P\leq70$~s) need to be re-evaluated. The observed power
density at high frequencies seems to be caused by line formation
effects in a dynamic atmosphere and cannot be interpreted as evidence
of propagating high frequency acoustic waves. Therefore, estimates of
the energy flux of high frequency acoustic waves should be treated
with caution, in particular those that apply atmospheric MTF
corrections.  On the other hand, narrowband filtergrams provide a
reasonable measure of the strong and dominant 3- and 5-min
oscillations if the filter position is chosen well (i.e. far from the
knee between the Doppler core and the Lorentzian damping wings).

\begin{acknowledgements}
The CO$^5$BOLD simulations were carried out at CINECA (Bologna/Italy)
with CPU time assigned under INAF/CINECA agreement 2006/2007. This
research was supported by the Research Council of Norway through
grants of computing time from the Programme for
Supercomputing. Th.S. acknowledges financial support by ESA and
ASI. SJ acknowledges financial support from NSF.  Hinode is a Japanese
mission developed and launched by ISAS/JAXA, with NAOJ as domestic
partner and NASA and STFC (UK) as international partners. It is
operated by these agencies in co-operation with ESA and NSC (Norway).
\end{acknowledgements}

\bibliographystyle{aa}

\begin{thebibliography}{6}
\expandafter\ifx\csname natexlab\endcsname\relax\def\natexlab#1{#1}\fi

\bibitem[{{Fleck} {et~al.}(2008){Fleck}, {Jefferies}, {McIntosh}, 
{Severino}, {Straus}, \& {Tarbell}}]{2008ESPM...12.2.39F}{Fleck}, B., 
{Jefferies}, S.~M., {McIntosh}, S.~W., {et~al.} 2008, 12th ESP Meeting, 
Freiburg, Germany, Online at http://espm.kis.uni-freiburg.de

\bibitem[{{Hansteen} {et~al.}(2007){Hansteen}, {Carlsson}, \& 
{Gudiksen}}]{2007ASPC..368..107H}{Hansteen}, V.~H., {Carlsson}, M., \& 
{Gudiksen}, B. 2007, in The Physics of Chromospheric Plasmas, 
ed. {P.~Heinzel, I.~Dorotovi\v{c}, \& R.~J.~Rutten}, PASPC, 368, 107

\bibitem[{{Keil} \& {Marmolino}(1986)}]{1986ApJ...310..912K}
{Keil}, S.~L. \& {Marmolino}, C. 1986, \apj, 310, 912

\bibitem[{{Souffrin}(1966)}]{1966AnAp...29...55S}
{Souffrin}, P. 1966, Annales d'Astrophysique, 29, 55

\bibitem[{{Straus} {et~al.}(2008){Straus}, {Fleck}, {Jefferies}, {Cauzzi},
  {McIntosh}, {Reardon}, {Severino}, \& {Steffen}}]{2008ApJ...681L.125S}
{Straus}, T., {Fleck}, B., {Jefferies}, S.~M., {et~al.} 2008, \apjl, 681, L125

\bibitem[{{Wedemeyer} {et~al.}(2004){Wedemeyer}, {Freytag}, {Steffen},
  {Ludwig}, \& {Holweger}}]{2004A&A...414.1121W}
{Wedemeyer}, S., {Freytag}, B., {Steffen}, M., {Ludwig}, H., \& {Holweger}, H.
  2004, \aap, 414, 1121

\end{thebibliography}

\end{document}